\documentstyle[12pt]{article}
\normalbaselines
\begin{document}

\textheight=22 true cm
\textwidth=15 true cm

%DEFINITIONS

\def\a {\alpha}
\def\as {\alpha_S}
\def\ast {\alpha_*}
\def\ast0 {\alpha_{*,0}}
\def\g {\gamma}
\def\G {\Gamma}
\def\pizz {\Pi_{ZZ}}
\def\piww {\Pi_{WW}}
\def\pigg {\Pi_{\gamma\gamma}}
\def\pigz {\Pi_{\gamma Z}}
\def\pitt {\Pi_{33}}
\def\pitq {\Pi_{3Q}}
\def\piqq {\Pi_{QQ}}

\def\p {\Pi}
\def\mzs {M_Z^2}
\def\mws {M_W^2}
\def\mzsz {{M_Z^0}^2}
\def\mwsz {{M_W^0}^2}
\def\q2 {q^2}
\def\zzs {Z_{Z^*}}
\def\zws {Z_{W^*}}
\def\d {\delta}
\def\sz {\sin^2\theta_W|_Z}
\def\cz {\cos^2\theta_W|_Z}
\def\sst {s_*^2}

\def\ts {\tilde {\rm S}}
\def\tt {\tilde {\rm T}}
\def\tu {\tilde {\rm U}}
\def\tv {\tilde {\rm V}}
\def\tw {\tilde {\rm W}}
\def\tx {\tilde {\rm X}}
\def\r {\rightarrow}
\def\t {\times }
\def\qm {q^2 M_Z^{-2}}
\def\mq {M_Z^2\over q^2}
\def\ms {10^{-3}S}
\def\mt {10^{-3}T}
\def\miu {10^{-3}U}
\def\mv {10^{-3}V}
\def\mw {10^{-3}W}
\def\mx {10^{-3}X}

\def\be {\begin{equation}}
\def\ee {\end{equation}}
\def\bea {\begin{eqnarray}}
\def\eea {\end{eqnarray}}

%THE TEXT STARTS HERE
\thispagestyle{empty}
%\rightline{\large\sf TIFR/TH/96-15}
%
%\rightline{\large\sf March 1996}
%
%\rightline{\bf Revised on: 1st June 1996}

\begin{center}
{\Large\bf
A General Treatment of Oblique Parameters}\\[20mm]
{\large{\sf Anirban Kundu 
\footnote{Electronic address: akundu@theory.tifr.res.in}}
and {\sf Probir Roy
\footnote{Electronic address: probir@theory.tifr.res.in}}
}\\[5mm]
{\em Theoretical Physics Group,\\
Tata Institute of Fundamental Research,\\
Homi Bhabha Road, Bombay - 400005, India.}
\end{center}
\begin{abstract}

A reexamination is made of one-loop oblique electroweak corrections.  General 
definitions are given of the oblique parameters without reference to any 
$q^2$-expansion scheme. The old oblique parameters S,T and U are defined as 
differences of gauge boson vacuum polarization $\Pi$-functions and suffice 
to describe certain observable ratios on the $Z$-peak and the $\rho$ parameter 
at $q^2=0$. Regarding the new oblique parameters V,W and X, the first two  
are defined in terms of differences of $\Pi$-functions as well as the
wavefunction renormalization of the corresponding weak boson, and the 
third in terms of the difference of differences of two $\Pi$-functions for 
$\gamma-Z$ mixing. Explicit expressions for measurable quantities involving 
all six oblique parameters are given and experimental bounds are obtained on
the latter, some for the first time. A review of these constraints suggests
that the linear approximation of Peskin and Takeuchi is robust.

\end{abstract}

\newpage

\centerline {\large I. {\bf Introduction}}

\bigskip

A lot of interest in recent years has centred around electroweak 
precision parameters \cite{one}. These are very useful in describing the 
possible effects of interactions {\em beyond the Standard Model} (BSM).
Hints for such interactions have been sought ardently in many precision
experiments. To this end, one has had to first carefully compute 
various quantities in the Standard Model (SM). These calculations take
into account electroweak one-loop radiative corrections \cite{two} as
well as higher loop corrections in QCD and those involving the top quark.
Given these computed results, tiny deviations between predicted and 
measured values are looked for. In their absence, bounds are put on the
precision parameters.

Consider processes without any external top quark. The coupling between
the $Z$ boson and a bottom quark-antiquark pair is special in that the
specific top-mediated large vertex correction needs to be explicitly 
included. Otherwise, to a reasonable accuracy, one-loop corrections 
to all remaining SM quantities, measured to date, can be approximated by
gauge vector-boson self-energy graphs \cite{pt}, known as {\em oblique
corrections.}
(For more precise estimates \cite{four}, of course, general vertex 
corrections and  box graphs at the one-loop level as well as top-induced
two-loop terms need to be taken into account.) The aforementioned parameters 
can now be defined as linear combinations \cite{five,six} of differences
between various gauge boson vacuum polarization $\p$-functions at the
same scale or at two separate scales and henceforth will be called
old oblique parameters.
These are the three constants S,T,U or equivalently $\epsilon_1$,
$\epsilon_2$,$\epsilon_3$ \cite{one} --- with $\epsilon_1=\a T$,
$\epsilon_2=-\a(4s^2)^{-1}U$,$\epsilon_3=\a(4s^2)^{-1}S$, $\a$ and $s$
respectively being the fine structure constant and the sine of the
Weinberg angle at the tree level.
LEP and lower energy data have been used to obtain interesting bounds
on them \cite{pt,four,five,six}.

One can classify the nongauge SM fields into light (all fermions except
the top) and heavy (top and Higgs) sectors. The contributions of the light
fields to perturbatively calculable quantities are known quite precisely,
those from the heavy sector less so. Moreover, BSM contributions --- if
any --- are likely to get mixed up with the latter. The effects of such
heavy physics can be added on to the light physics SM calculation by using
the $\p$-formalism. We define $\p_{AB}(q^2)$ in terms of the real part
of the current propagator 
\be
{\rm Re}~\int~d^4xe^{iq.x}\langle \Omega|T J_\mu^A(x)J_\nu^B(0)|\Omega\rangle
=-\p_{AB}(q^2)\eta_{\mu\nu}+q_\mu q_\nu~{\rm terms},
\ee
$\eta_{\mu\nu}$ being the flat Minkowski metric and A,B being gauge group
indices. The $q_{\mu}q_{\nu}$ terms drop out when contracted with the 
external fermion current (as we are considering light fermions only in
the external legs), so that $\Pi_{\mu\nu}$, and all quantities derived
from it, are gauge invariant. 
Physics enters the LHS of (1) through the insertion of  a 
complete set of states $\sum_n|n\rangle\langle n|$ between the current
operators. The sum can be decomposed neatly as $\sum_{n_l}|n_l\rangle\langle
n_l|$ $+$
$\sum_{n_h}|n_h\rangle\langle
n_h|$ where $|n_l\rangle$ refers to the light SM sector 
and $|n_h\rangle$ to the heavy sector including both heavy SM and any possible
BSM fields. Thus the heavy and the light sector contributions add 
linearly to the $\p$-functions and hence to the oblique parameters. 

One can also consider the difference of two $\p$-functions at different 
$q^2$-values by defining a function $\p'_{AB}(q^2)$:
\be
\p_{AB}(q^2)\equiv\p_{AB}(0)+q^2\p'_{AB}(q^2).
\ee
At $q^2=0$, $\p'_{AB}(q^2)$ and $d\p_{AB}(q^2)/dq^2$ are equal; otherwise
they are different. This procedure can be extended one step further 
by introducing $\p''_{AB}(q^2)$. This is defined as proportional to
the difference of two $\p'_{AB}$ functions, {\em i.e.}, to the 
difference of differences of $\p_{AB}$ functions. Thus define
\be
\p''_{AB}(q^2)\equiv 2q^{-2}[\p'_{AB}(q^2)-\p'_{AB}(0)].
\ee
Again, this equals $d^2\p_{AB}(q^2)/d(q^2)^2$ only at $q^2=0$. It
follows from (2) and (3) that 
\be
\p_{AB}(q^2)=\p_{AB}(0)+q^2\p'_{AB}(0)+{1\over 2}q^4\p''_{AB}(q^2).
\ee

The extraction of S,T,U from the data, obtained in experiments performed
both on the $Z$- and the $W$-masses as well as at low energies, has so far
been made with some kind of an approximation on the $q^2$-dependence of 
the concerned $\p$-functions. Peskin and Takeuchi proposed \cite{pt} 
a linear expansion approximation which we briefly recount. This approximation
amounts to replacing $\p'(q^2)$ by $\p'(0)$. Thus (2) becomes
\be
\p_{AB}(q^2)=\p_{AB}(0)+q^2\p'_{AB}(0).
\ee
A quadratic expansion approximation, which is a quadratic extension of
(5), {\em i.e.},
\be
\p_{AB}(q^2)=\p_{AB}(0)+q^2\p'_{AB}(0)+{q^4\over 2}\p''_{AB}(0),
\ee
has also been suggested in \cite{seven}. Our aim in this paper 
is to avoid --- to the extent possible --- this kind of approximation 
in extracting the oblique parameters from experimental data. 
Thus, we need to identify those segments of the data which make
this feasible for S,T and U. We will see that the utilization
of the full set of electroweak data in determining the oblique
parameters requires the introduction of three more parameters \cite{eight}
V,W and X. These vanish in the linear approximation and 
will be called the new oblique parameters.

Before proceeding further, it is useful to review the motivation behind the
linear approximation employed in  \cite{pt} and related subsequent 
works \cite{five,six}. In the SM the difference between (2) and (5) 
for $0\ll q^2\ll M_{W,Z}^2$ can be computed and is known to be negligibly tiny.
Coming to BSM contributions, Peskin and Takeuchi \cite{pt} were inspired
by scenarios of electroweak symmetry breakdown, in which the BSM mass scale M 
controlling the dimensional denominators in $\p'(0)$, $\p''(0)$ etc. was
expected to be ${>\atop\sim}$1 TeV. In such a situation, since $q^2$
varies between zero and $M_{W,Z}^2$ only, it would not make much sense
to retain the ${\cal O}(q^4/M^4)$ terms, specially when one is neglecting
both the non-oblique corections to one-loop terms and the two-loop 
contributions. In this picture the linear approximation could a priori be
deemed as accurate. There are, however, other scenarios of electroweak 
symmetry breaking in which some BSM scales could have a significantly
lower value --- such as 100-200 GeV. An example would be a low-lying 
techniparticle or a low electroweak gaugino mass in supersymmetry 
\cite{nine}. Now the accuracy of the linear approximation could be called
into question. It would be desirable to give more general definitions
of the oblique parameters, independent of any $q^2$-expansion, and hence 
without any commitment to the magnitude of the scale controlling
electroweak symmetry breaking dynamics.

In this paper we formulate these general definitions. The old oblique 
parameters S,T,U are expressed as differences \cite{six} of $\Pi$-functions
while the new ones V,W and X are given in terms of corresponding differences
of differences and their derivatives. A similar approach was taken earlier
\cite{eight}, but we have made several improvements. We give definitions
that preserve the correct symmetry properties of S and U unlike in
\cite{eight}. Furthermore, not only do we describe all one-loop oblique
electroweak corrections in terms of the full set of six parameters
S,T,U,V,W and X, we are able to put bounds on {\em all} of them using the
data. Moreover, we clarify the issue of $q^2$-expansion and discuss 
the robustness of the linear approximation.  

Focussing on the $q^2$-expansion procedure, we examine the question why
it was used by previous authors 
\cite{pt,five,seven} at all. This expansion entered the
picture only in relating the $Z$- and $W$-wavefunction renormalization
constants which involve derivatives of $\p$-functions to S,T,U which
involve only differences. That step was necessitated in the calculation of 
certain physical observables in terms of those parameters since the
said renormalization constants occur in their expressions. Ratios, which
do not depend upon the wavefunction renormalization constants, are
directly computable in terms of S,T and U without any use of the 
expansion approximation. Turning then to quantities which do, we show 
precisely how V,W and X can be defined so that all such quantities 
are expressible in terms of six oblique parameters. Using the first set
of ratios and experimental data, one can obtain numerical bounds on S,T 
and U {\em independently of any expansion approximation}. These can then be
utilized with respect to experimental data on the second set of quantities 
to derive the best bounds on V,W and X. The bounds on V and X are again
obtained without any expansion; however, that on W is derived within the 
quadratic approximation. SM contributions to all the oblique 
parameters (with a reference point of top and Higgs mass values) can then be
subtracted to yield bounds on the corresponding BSM contributions.

The rest of the paper is organized as follows. Section II contains a review
of the definitions of the old oblique parameters S,T and U, emphasizing 
their symmetry content. Section III discusses the issue of wavefunction 
renormalization of the weak vector bosons and the two new oblique parameters 
V,W as well as the need for the sixth parameter X; general definitions of 
these three are also given. In Section IV explicit one-loop corrected 
formulae (listed in the Appendix) for various observables are derived 
in the $\star$ scheme in terms of all six oblique parameters distinguishing
between certain ratios which involve only S,T and U, and quantities which
also involve the new parameters. Numerical constraints on the old and
new oblique parameters are discussed in Section V. The final Section VI
summarizes our conclusions and contains a discussion of the robustness 
of the linear approximation. 

\bigskip\bigskip

\centerline{\large II. {\bf Old Oblique Parameters}}

\bigskip

Though a sizable literature exists on S,T and U, it would be useful to
review their definitions here before introducing the new oblique
parameters. In the generic $\p$-function $\p_{AB}(q^2)$ of (1), the
combination (AB) is allowed to take the values (11), (22), (33), (3Q) 
and (QQ). Here 1,2,3 refer to the generators of the $SU(2)_L$ gauge
group and Q to the electromagnetic current. On account of the unbroken
part of the weak isospin symmetry, $\p_{11}$ equals $\p_{22}$; hence
there are four independent $\p$-functions. These are generally written
in the gauge boson basis as follows:
\bea
\pigg&=&e^2\piqq ,\\
\pigz&=&{e^2\over sc}(\pitq-s^2\piqq) ,\\
\pizz&=& {e^2\over s^2c^2}(\pitt-2s^2\pitq+s^4\piqq) ,\\
\piww&=&{e^2\over s^2}\Pi_{11}.
\eea 
In our notation the electromagnetic, the weak charged and the weak neutral
current interaction terms in the Lagrangian density are given respectively
by
\bea
{\cal L}_{em}&\equiv&eJ_\mu^QA^\mu ,\\
{\cal L}_{W;cc}&\equiv&{e\over\sqrt{2}s}(J_{\mu}^{1-i2}W^{\mu +}
+J_\mu^{1+i2}W^{\mu -}) , \\
{\cal L}_{W;nc}&\equiv&{e\over sc}(J_\mu^3-s^2J_\mu^Q)Z^\mu ,
\eea
with $A^\mu$, $W^{\mu\pm}$ and $Z^\mu$ as the corresponding electromagnetic,
charged weak and neutral weak gauge boson fields respectively. The tree-level
quantities $s^2$ and $e\equiv\sqrt{4\pi\a}$ are defined from (i) $s^2=
1-\mwsz/\mzsz$ where $M_W^0$ and $M_Z^0$ are the tree-level masses of the
$W$ and the $Z$ respectively (without the finite self-energy corrections)
and (ii) the Thompson cross-section $\sigma_T=8\pi(\a/m_e)^{2/3}$ \cite{ten}.
The tree-level four-Fermi coupling $G_F^0$ is given by $\sqrt{2}e^2/
(8s^2c^2\mzsz)$.

Let us restrict ourselves only to on-shell or near-peak $Z$,$W$ data plus
low energy measurements which define the electroweak couplings. Let us
also first consider dimensionless ratios of measurable quantities from which 
the $W$ or $Z$ wavefunction renormalizations cancel out. Then it suffices 
to take $\pigg$, $\pigz$ and $\pizz$ at $q^2=\mzs,0$ and $\piww$ at
$q^2=\mws,0$. That would give us eight one-loop corrected quantities, but
the QED Ward identities imply
\be
\pigg(0)=\pigz(0)=0,
\ee
leaving six such independent quantities. We have already related tree-level
parameters directly to experimental data. For one-loop corrected quantities,
one would like to utilize three accurate numbers obtained from three 
precision measurements \cite{ten}: (i) the inverse of the fine structure
constant, measured \cite{pdg} 
from the Quantum Hall Effect, to be 137.036, (ii) the
mass of the $Z$ determined from LEP --- namely (91.1888$\pm$ 0.0022) GeV,
and (iii) the Fermi constant, inferred from the muon lifetime calculated 
to one-loop, as $1.16639(2)\times 10^{-5}$ GeV$^{-2}$. These can be used
to eliminate three of the $\p$s. Finally, three more one-loop corrected
quantities remain and the system is completely described by the choice
\cite{six}: 
\be
S\equiv -{8\pi\over \mzs}\Big[\p_{3Y}(\mzs)-\p_{3Y}(0)\Big],
\ee
\be
T\equiv {4\pi\over c^2s^2\mzs}\Big[\p_{11}(0)-\p_{33}(0)\Big],
\ee
\be
U\equiv{16\pi\over \mws}\Big[\Pi_{11}(\mws)-\p_{11}(0)\Big] 
-{16\pi\over \mzs}\Big[\Pi_{33}(\mzs)-\p_{33}(0)\Big].
\ee
In (15) $\p_{3Y}=2(\p_{3Q}-\p_{33})$ since $J_\mu^Q=J_\mu^3+\textstyle
{1\over 2}J_\mu^Y$.

What needs to be stressed in the definitions (15)-(17) is the physical
meaning behind the symmetry content of each of the three oblique 
parameters. S quantifies the difference in mixing between the 
hypercharge and the third weak isospin currents at $q^2=\mzs$ and 
$q^2=0$ (the mixing itself between two operators in the two factor groups 
is absent classically but arises at the quantum one-loop level owing to
spontaneous symmetry breakdown). In contrast, T and U are directly 
concerned with weak isospin rather than hypercharge. T describes the
amount of weak isospin breaking at $q^2=0$ and is linearly proportional
to the deviation from unity of the $\rho$-parameter, measured at low energies. 
On the other hand, U measures the contribution of the $W$,$Z$ mass 
nondegeneracy to weak isospin breaking. The additional oblique parameters 
V,W and X get introduced when one goes beyond measurable quantities 
which do not depend on the wavefunction renormalization and starts to
consider those which do. Nevertheless, the definitions (15)-(17) stand and
do not suffer any modification since the latter are independent of the 
$q^2$-expansion procedure.

At this point, it would be useful to contrast our definitions (15)-(17)
with those of ref. \cite{eight}. We have extended the S,T,U definitions of
Peskin and Takeuchi (PT) \cite{pt} beyond the linear approximation to 
a $q^2$-independent form. Burgess {\em et al} \cite{eight} aimed to do
the same with the Marciano-Rosner (MR) \cite{five} definitions of the 
corresponding quantities. With only linear terms retained, the PT and the
MR definitions agree while ours agree with those of ref. \cite{eight}
only for T, but not for S and U. The difference concerns the broken and
custodial symmetry contents of the $\p$-function combinations relevant 
to these oblique parameters. Specifically, we can show that while our 
definitions preserve the desired symmetry properties, theirs do not. One
could, of course, argue that the choice of definitions is a matter of
convenience since it is possible to go from phenomenological fits for
one set to those for the other by appropriate substitutions. However, 
different definitions mean different physical quantities and this 
important point needs to be kept in mind.

Let us reconsider the definitions of S and U. Recasting (15) and (17) 
in the gauge boson basis [cf. (7)-(10)], we have
\be
{\a\over 4s^2c^2}S={\pizz(\mzs)-\pizz(0)\over \mzs}-{c^2-s^2\over cs}
{\pigz(\mzs)\over\mzs}-{\pigg(\mzs)\over\mzs},
\ee
\bea
{\a\over 4s^2}U={\piww(\mws)-\piww(0)\over\mws}&-& c^2{\pizz(\mzs)
-\pizz(0)\over\mzs}\nonumber\\
&{ }&-s^2{\pigg(\mzs)\over\mzs}-2sc{\pigz(\mzs)\over\mzs}.
\eea
In place of the ratios $\pigz(\mzs)/\mzs$ and $\pigg(\mzs)/\mzs$, the 
authors of \cite{eight} had used $\p'_{\g Z}(0)$ and $\p'_{\g\g}(0)$
respectively. With such a choice, S could not be written as a difference
of $\p_{3Y}$ functions between the scales $q^2=\mzs$ and $q^2=0$ except
only in the linear approximation. Thus the physical meaning of S as a 
measure of the quantum mixing between $J_\mu^3$ and $J_\mu^Y$ currents
would be lost. Similar would be the fate of U as a measure of the 
contribution of the $W$,$Z$ mass nondegeneracy to weak isospin breaking. 
The ``S" and ``U", defined in \cite{eight}, are therefore quantities that
are physically different from what have been considered so far. The two
sets are related, via the X-parameter (see Section III), by ``S"=S$-4(c^2
-s^2)$X, ``U"=U$-8s^2$X. In consequence, renormalization independent 
observable ratios, which depend only on S,T and U, become functions of
``S",T,``U" and X.

An alternative approach might be to replace, in the above definitions, 
the differences $[\pizz(\mzs)-\pizz(0)]/\mzs$ and $[\piww(\mws)
-\piww(0)]/\mws$ by $\p'_{ZZ}(0)$ and $\p'_{WW}(0)$ respectively, 
alongwith the replacements done in ref. \cite{eight}. Now, `S'=$-8
\pi\p'_{3Y}(0)$ and `U'=$16\pi[\p'_{11}(0)-\p'_{33}(0)]$, so that
`S' still quantifies the 3Y mixing albeit at $q^2=0$, while `U' measures
weak isospin breaking for the $\p'(0)$s. With these definitions, however,
no physical observable involving `S' and `U' at the one-loop level would
be computable without invoking the $q^2$-expansion procedure. Our definitions
(15)-(17) have been chosen so as to preserve the symmetry properties of the
oblique parameters {\em as well as} to enable the determination of
S,T,U without reference to that expansion approximation.

\bigskip\bigskip

\centerline{\large III. {\bf Renormalization and the New Oblique
Parameters}}

\bigskip

Once again, it would be useful to start with a brief review -- this
time of the wavefunction renormalizations of the $Z$
and of the $W$ and of their role in computing one-loop observables --
along the lines of Peskin and Takeuchi \cite{pt}.  The tree-level masses of
the gauge bosons $M_W^0$, $M_Z^0$ get modified by
vacuum polarization diagrams.  Now the 
wavefunction renormalization constants $Z_V$ (V=W,Z) ---
defined consistently to one-loop as the coefficients of the poles
of the corresponding propagators  
$Z_V\equiv (1-{d/dq^2}~\Pi_{VV}|_{q^2=M_V^2})^{-1}$
$\approx 1+{d/dq^2}\Pi_{VV}|_{q^2=M_V^2}$ ---  develop
nontrivial dependences on the oblique parameters. 
By using (4), one can then write
\be
Z_Z=1+\Pi'_{ZZ}(0)+M_Z^2\Pi''_{ZZ}(M_Z^2)+{1\over 2}M_Z^4{d\over dq^2}
\Pi''(q^2)|_{q^2=M_Z^2}, \ee
\be
Z_W=1+\Pi'_{WW}(0)+M_W^2\Pi''_{WW}(M_W^2)+M_W^4{d\over dq^2}
\Pi''(q^2)|_{q^2=M_W^2}. \ee
If the concerned $\Pi$-functions were only linearly dependent on $q^2$
\cite{pt}, $d~/dq^2~\Pi|_{q^2=M_V^2}$ would be expressible 
as proportional to the
difference $\Pi(M_V^2)-\Pi(0)$ and would not be an independent quantity.
However, any general nonlinear $q^2$-dependence in $\Pi_{VV}(q^2)$ would
make the derivative independent of $\Pi(M_V^2)$ and $\Pi(0)$.

We now define two new oblique parameters V and W as follows:
\bea
\alpha V&\equiv& {1\over 2}\Big[ 
{d\over dq^2}[q^4\Pi''_{ZZ}(q^2)]_{q^2=M_Z^2}
-M_Z^2\Pi''_{ZZ}(q^2)\Big], \\
\alpha W&\equiv& {1\over 2}\Big[
{d\over dq^2}[q^4\Pi''_{WW}(q^2)]_{q^2=M_W^2}
-M_W^2\Pi''_{WW}(q^2)\Big]. \eea
These are new in the sense that they do not contribute in the linear
approximation \cite{pt} under which $\Pi''_{AB}(q^2)\to 0$. Further, under
the quadratic approximation \cite{seven}, 
namely $\Pi''_{AB}(q^2)\approx\Pi''_{AB}
(0)$, they become $\alpha V\approx \Pi'_{ZZ}(M_Z^2)-\Pi'_{ZZ}(0)$,
$\alpha W\approx \Pi'_{WW}(M_W^2)-\Pi'_{WW}(0)$.
It is to be noted that (20)-(23) yield the
$Z$, $W$ wavefunction renormalization constants to be
\bea
Z_Z&=& 1+M_Z^{-2}[\Pi_{ZZ}(M_Z^2)-\Pi_{ZZ}(0)]+\alpha V, \\
Z_W&=& 1+M_W^{-2}[\Pi_{WW}(M_W^2)-\Pi_{WW}(0)]+\alpha W. \eea
When we extend our consideration of measurable quantities on each vector 
boson pole from only dimensionless ratios (described by $\alpha$,$G_F$,$M_Z$
and S,T,U) to also include dimensional quantities, two more parameters,
namely, $Z_Z$ and $Z_W$ (equivalently two
new oblique parameters V and W), enter their formulae
calculated to one-loop.

In order to avoid the
$q^2$-expansion and analytically use the new
oblique parameters, we find it convenient to use
the $\star$ renormalization scheme of Kennedy and Lynn \cite{eleven}.  This is
for illustrative purposes only.  The $\star$ scheme has deficiencies
with respect to the level of precision now achievable in electroweak
testing.  It does not incorporate
the one-loop vertex corrections and box graphs
which contribute nonnegligibly to low-energy phenomena.  With the
leading log approximation, it also cannot fully cover significant
top-induced two-loop terms which are part and parcel of the present day
high precision calculations.  However, the focus in this paper is not on
the numerical accuracy of radiative corrections, but rather on a formal
discussion of $\Pi$-functions with analytic expressions. It is for that
purpose that we employ the $\star$ scheme.

In the $\star$ scheme, the key step is to replace all the bare parameters
at the tree-level with their starred counterparts
{\em evaluated at appropriate momenta}. Thus in this
scheme the
running wavefunction renormalizations $Z_{Z^\star} (q^2)$ and
$Z_{W^\star} (q^2)$ are given by \cite{pt} the relations
\be
\displaystyle{e^2_\star \over s^2_\star c^2_\star} Z_{Z^\star} =
\displaystyle{e^2 \over s^2 c^2} Z_Z, 
\ee
\be
\displaystyle {e^2 \over s^2_\star} Z_{W^\star} =
\displaystyle{e^2 \over s^2} Z_W. 
\ee
The quantities $e^2_\star (q^2)$, $s^2_\star (q^2)$ and $c^2_\star
(q^2)$, which appear in (26)-(27), are related at the one-loop level
to $e^2$, $s^2$ and $c^2$
respectively through appropriate $\Pi'$-functions as follows:
\bea
e^2_\star (q^2)& =& e^2 [1 + \Pi'_{\gamma\gamma} (q^2)], \\
s^2_\star (q^2)& =& s^2 \left[1 -  \displaystyle{c \over s}
\Pi'_{\gamma Z} (q^2)\right], \\
c^2_\star (q^2)& =& c^2 \left[1 + \displaystyle{s \over c}
\Pi'_{\gamma Z} (q^2)\right], 
\eea
and we maintain $s_*^2(q^2)+c_*^2(q^2)=1$. In (28)-(30)
we have once again retained only the terms that are linear in the
$\Pi'$s as a consistent one-loop approximation.
It now follows that
\bea
Z_{Z^*}(M_Z^2)&=&Z_Z{e^2\over s^2c^2}\Bigg({s_*^2c_*^2
\over e_*^2}\Bigg)_{M_Z^2}\\
&=&1+{\alpha\over 4s^2c^2}S+\alpha V, \eea
where we have used (18), (24), (26) and (28)-(30). Similarly,
\be
Z_{W^*}(M_W^2)=
1+{\alpha\over 4s^2}(S+U)+\alpha W. \ee
 
When we extend our set of calculable quantities to include measurables 
at $q^2=0$, two additional quantities enter, namely, 
$e_*^2(0)$ and $s_*^2(0)$. Of these, $e_*^2(0)$ equals $4\pi\alpha_*(0)$
where $\alpha_*(0)$ is the fine structure constant measured \cite{pdg}
from the
Quantum Hall Effect. This has already been taken as an input parameter
(see Section II). In contrast, $s_*^2(0)$ is an independent quantity. 
While $s_*^2(M_Z^2)$, a dimensionless measurable on the $Z$-pole,
is given in terms of old oblique parameters, $s_*^2(0)$
can only be related to $s_*^2(M_Z^2)$ through the introduction of another 
new oblique parameter X to describe the running:
\be
\alpha X\equiv -sc[\Pi'_{\gamma Z}(M_Z^2)-\Pi'_{\gamma Z}(0)]. \ee
$\Pi'_{\gamma Z}(M_Z^2)$ has already appeared in S (vide eq. (18)), but 
$\Pi'_{\gamma Z}(0)$ is an independent quantity, not necessarily small. 
For example, in the SM $\p'_{\g Z}(0)$ is ${\cal O}(10^{-4})$. This 
underscores the need for the new parameter X.
By comparison, $\Pi'_{\gamma\gamma}(0)
\equiv 1-\alpha\alpha^{-1}_*(0)$; taking $\alpha$ from Thompson
scattering [10] and $\alpha_*(0)$ from the Quantum Hall effect \cite{pdg},
we find $\Pi'_{\gamma\gamma}(0)\approx 2\times 10^{-6}$, which can be
safely neglected. Henceforth we shall therefore
replace $\alpha_*(0)$ by $\alpha$
everywhere.  Returning to X, we see that it
vanishes in the linear approximation and that its definition (34)
agrees with that given in ref. \cite{eight}. 
An additional point to note is that
neither $e_*^2(M_W^2)$ nor $e_*^2(M_Z^2)$
is a measurable quantity on the $W$- or $Z$-pole.
In order to make contact with 
experimentally measured dimensional quantities on each vector boson pole,
one needs to theoretically evolve $e_*^2$ from $q^2=0$ to $q^2=M_W^2,M_Z^2$,
via the renormalization group,
calculating all the light SM contributions to it ({\em i.e.}, 
excluding those of the top and the Higgs). This quantity will be
called \cite{pt} $e_{*,0}^2(M_V^2)$, and we shall use
$\alpha_{*,0}(M_V^2)\equiv e_{*,0}^2(M_V^2)/(4 \pi)$, (V=W,Z).

\bigskip\bigskip
\newpage

\centerline {\large IV. {\bf Observables in Terms of Oblique
Parameters}}

\bigskip

We now present the calculations of several observables and
check their dependences on the oblique parameters, old and new.
First, let us consider the $\rho$-parameter which involves the ratio of
the neutral current neutrino scattering amplitude to the charged
current one at $q^2=0$.  The $q^2=0$ condition can be used to one's
advantage by writing \cite{pt}
\be
\rho_\star (0) = 1 + {4\pi\alpha \over s^2c^2M^2_Z} [\Pi_{11} (0) -
\Pi_{33} (0)] = 1 + \alpha T.
\ee
It may be noted here that in the low-energy limit the one-loop corrected
four fermion weak interaction term is
\be
\displaystyle{G_F \over \sqrt{2}} \left[J^+_\mu J^{\mu^-} + \rho_\star
(0) J_\mu^{NC} J^{\mu NC}\right],
\ee
where $J^+_\mu \equiv J_\mu^1-iJ_\mu^2$ and $J_\mu^{NC} \equiv
J_\mu^3 - s^2_\star (0) J^Q_\mu$.  
We ought to reemphasize here that in terms of numerical accuracy that is
now attainable \cite{four} in precision testing at $q^2=0$, the obliqueness 
approximation and hence the $\star$ formalism is rather inaccurate since
contributions from non-oblique terms are not entirely negligible.

We come next to $s_*^2$. At the tree-level, we have
\be
s^2c^2={\pi\a\over\sqrt{2}G_F^0M_Z^{0^2}}.
\ee
Our $Z$-standard \cite{pt} is to identify $\sz$, defined by
\be
\sz\equiv{1\over 2}\Bigg[1-\Big\{1-{4\pi\a_{*,0}(\mzs)\over\sqrt{2}
G_F\mzs}\Big\}^{1/2}\Bigg],
\ee
as the renormalized value of $s^2$, {\em i.e.}, equal to
\be
s^2+\d s^2=s^2+{s^2c^2\over c^2-s^2}\d\ln(s^2c^2).
\ee
with $\alpha$ evolved to $\alpha_{0,\star} (M^2_Z)$.  (37) then leads to
\be
\sz-s^2={s^2c^2\over c^2-s^2}\Bigg({\delta \alpha_{0,\star} \over
\alpha_{0,\star}}-{\d G_F\over G_F}
-{\d M^2_Z\over M^2_Z}\Bigg).
\ee
Following Lynn, Peskin and Stuart \cite{lynn}, we can write
\bea
{\d\alpha_{0,\star}\over\alpha_{0,\star}}&=&\p'_{\g\g}(M^2_Z),\\
{\d G_F\over G_F}&=& -{\piww(0)\over\mws},\\
{\d M^2_Z\over M^2_Z}&=&{\pizz(\mzs)\over\mzs}.
\eea
(40)-(43) yield
\bea
\sz-s^2&=&e^2{s^2c^2\over c^2-s^2}\Big[\p'_{QQ}(M^2_Z)+{1\over s^2c^2\mzs}\{
\Pi_{11}(0)-\Pi_{33}(\mzs)\nonumber\\
&{ }& ~~~~~~~~~~~~~~+2s^2\Pi_{3Q}(\mzs)-s^4\piqq(\mzs)\}\Big].
\eea
On the other hand, from (29), we can write
\be
s_*^2(q^2)-s^2=-e^2[\p'_{3Q}(q^2)- s^2 \p'_{QQ}(q^2)].
\ee
(44) and (45) lead to the results
\be
s_*^2(\mzs)=\sz+{\a\over c^2-s^2}({1\over 4}S-s^2c^2T),
\ee
\be
s_*^2(0)=\sz+{\a\over c^2-s^2}({1\over 4}S-s^2c^2T)-\a X.
\ee
We note that X does not appear in the observables measured on the $Z$-pole,
while it would appear in the low-energy observables that contain $s_*^2(0)$.
It is also noteworthy that we can write $s_*^2$ exactly only
at $q^2=0$ and $q^2=\mzs$ and at no other value of $q^2$, as is evident from
the expression for X. Thus, for example, in order to evaluate $s_*^2(\mws)$,
one has to use the quadratic expansion approximation \cite{seven}. This is
why the latter will be needed in obtaining bounds on W. It is also 
interesting that X appears through $s_*^2(0)$ in the expression 
\cite{czarnecki} for the left-right parity-violating Moller asymmetry.

The ratio of the gauge boson masses can be expressed in terms of the old
oblique parameters. Starting with
\bea
\mws&=&\mwsz+{e^2\over s^2}\p_{11}(\mws)\nonumber\\
&=&(\mzs-\d\mzs)(\cz-\d\cz)+
{e^2\over s^2}\p_{11}(\mws),
\eea
and using (40), we derive
\be
M^2_W/M^2_Z = \cos^2 \theta_W|_Z + {\alpha c^2 \over c^2 - s^2}
\Bigg[ -{1\over2} S + c^2T 
+ {c^2 -s^2 \over 4s^2} U \Bigg].
\ee
(The derivation proceeds much in the same manner as that given in the
longer article by Peskin and Takeuchi \cite{pt}.)

The partial widths measured at the $Z$-pole contain S and T among the 
old obique parameters, and V among the new ones. This is evident from the
following expressions in the $\star$ scheme.
\be
\Gamma (Z \rightarrow \nu\bar \nu) = Z_{Z^\star} (M^2_Z) {\alpha_{\star,0}
(M^2_Z) M_Z \over 24 s^2_\star (M^2_Z) c^2_\star (M^2_Z)},
\ee
\bea
\Gamma (Z \rightarrow \ell\bar\ell) &=& Z_{Z^\star} (M^2_Z)
\displaystyle{\alpha_{\star,0} (M^2_Z) M_Z \over 6s^2_\star (M^2_Z)
c^2_\star (M^2_Z)} (1 + \delta_{\rm QED}) \nonumber\\
& { } & \left[\left\{-\displaystyle{1\over2} + s^2_\star
(M^2_Z)\right\}^2 + \left\{s^2_\star (M^2_Z)\right\}^2\right],
\eea
\bea
\Gamma (Z \rightarrow u\bar u) &= &\Gamma (Z \rightarrow c\bar c) =
Z_{Z^\star} (M^2_Z) \displaystyle{\alpha_{\star,0} (M^2_Z) M_Z \over
2s^2_\star (M^2_Z) 
c^2_\star (M^2_Z)} (1 + \delta_{\rm QED} + \delta_{\rm QCD})\nonumber \\
&{ }&\left[\left\{\displaystyle{1\over2} - \displaystyle{2\over3}
s^2_\star (M^2_Z)\right\}^2 + 
\left\{-\displaystyle{2\over3} s^2_\star (M^2_Z)\right\}^2\right],
\eea
\bea
\Gamma (Z \rightarrow d\bar d) &= &\Gamma (Z \rightarrow s\bar s) =
Z_{Z^\star} (M^2_Z) \displaystyle{\alpha_{\star,0} (M^2_Z) M_Z \over
2s^2_\star (M^2_Z) 
c^2_\star (M^2_Z)} (1 + \delta_{\rm QED} + \delta_{\rm QCD}) \nonumber\\
&{ }&\left[\left\{-\displaystyle{1\over2} + \displaystyle{1\over3}
s^2_\star (M^2_Z)\right\}^2 + 
\left\{\displaystyle{1\over3} s^2_\star (M^2_Z)\right\}^2\right],
\eea
where 
\be
\delta_{\rm QED} = 3\a Q^2_f/4\pi, \ \ 
\delta_{\rm QCD} 
= {\as \over \pi} + 1.409 \left({\alpha_S \over \pi}\right)^2
- 12.77 \left({\alpha_S \over \pi}\right)^3,
\ee
$\alpha_S$ being the QCD fine structure constant and a colour factor of
3 has been included in the $Z\to q\bar q$ cases. The decay of $Z$ into a 
$b\bar b$ pair is on a slightly different footing, since the vertex 
correction mediated by a virtual top quark has to be taken care of. This
correction can be accounted for by writing $\Delta\rho_t = 3m_t^2G_F/
(8\pi\sqrt{2})$ and replacing $s_*^2$ in (53) by $s_*^2(1+\textstyle
{2\over 3}\Delta\rho_t)$, assuming that the BSM physics does not contribute
significantly to the non-oblique part.

Employing the numerical values of $\a$, $\a_{*,0}(\mzs)$, $\as$, $M_Z$ and
$G_F$, one can write these $Z$-pole observables, as well as the total decay
width $\Gamma_Z$, as functions of S,T and V, as shown in the Appendix. 
In contrast, the ratio of any two partial widths of the $Z$ --- {\em
e.g.}, $\G(Z\to {\rm hadrons})/\G(Z\to \ell\bar \ell)$ --- does not contain
V, which is also true for all $Z$-pole asymmetries. Thus, these ratios
and asymmetries put bounds on S and T, while the absolute values of the
total and partial decay widths of $Z$ contain V also. Now one can use eq. (49)
to put a bound on U. This is somewhat of a loose bound at the moment, but 
once $M_W$ is known with an accuracy comparable to that of $M_Z$ --- from the
Tevatron and from LEP-II --- the bound on U would become quite stringent.

We now consider the $W$-width, which is also expected to be measured quite
accurately at LEP-II. The one-loop corrected
leptonic decay width of the $W$ is given in the
$\star$ scheme by
\be
\Gamma^\ell_W\equiv
\Gamma (W \rightarrow \ell\bar\nu) = Z_{W^\star} (M^2_W) {\alpha_{\star,0}
(M^2_W) (M_W/M_Z) \over 12s^2_\star (M^2_W)} M_Z.
\ee
The corresponding total width is
\bea
\Gamma_W &=& 3\Gamma_W^\ell + 3\{1 + \delta_{\rm QCD}\}
\Gamma_W^\ell \{|V_{ud}|^2 + |V_{cd}|^2 + |V_{us}|^2 + |V_{cs}|^2\}
\nonumber\\
&=& [3 + 3.123(1 + |V_{tb}|^2)]\Gamma_W^\ell.
\eea
$\a_{*,0}(\mws)$ comes out as $128.89\pm 0.12$ from logarithmic
running. Further, for the evolution from $\mzs$ to $\mws$, the use
of the quadratic approximation \cite{seven} may not be too bad. This yields
\be
s^2_\star (M^2_W) = \sin^2 \theta_W\big|_Z + {\alpha \over c^2 - s^2}
\left({1\over4} S - c^2s^2 T\right) - \alpha s^2 X.
\ee
The expressions for the total and the partial widths of the $W$, as shown
in the Appendix, contain S,T,U,W and X.

Finally, we come to observables related to low-$q^2$ experiments.
For deep inelastic neutrino scattering the quark couplings are 
\be
g^2_L \equiv [\rho_\star (0)]^2 [(g^u_{L\star})^2 + (g^d_{L\star})^2],
\ee
\be
g^2_R \equiv [\rho_\star (0)]^2 [(g^u_{R\star})^2 + (g^d_{R\star})^2].
\ee
The ratios of the neutral to charged current cross sections in the
neutrino and antineutrino cases can be written in terms of the
above couplings as
\be
R_\nu = g^2_L + r^0 g^2_R,
\ee
\be
R_{\bar\nu} = g^2_L + \displaystyle{1 \over \bar r^0} g^2_R,
\ee
where the numerical parameters $r^0$ and $\bar r^0$ are
$0.383 \pm 0.014$ and $0.371 \pm 0.014$ respectively.  Explicit 
expressions for $g^2_{LR}$ and $R_{\nu,\bar\nu}$, in terms of the
oblique parameters $S,T$ and $X$, appear in the Appendix.  Last, but not the
least, is the atomic weak charge of cesium which is determined by the
parity-violation experiment on cesium vapour:
\be
Q_W (^{133}_{55}Cs) = -\rho_\star (0) [78 - (1 - 4s^2_\star (0))55].
\ee
Like other experimental quantities measured at low energies, this
also is a linear combination of S,T and X, as given in the
last line of the Appendix.

\bigskip\bigskip

\centerline{\large V. {\bf Constraints on the New Oblique Parameters}}

\bigskip

One can put constraints on the oblique parameters, both old and new,
as follows. Compare the analytical expressions shown in the previous
section, and their numerical counterparts given in the Appendix, with
the experimental data. First, choose the data for only those quantities
whose expressions contain S,T and U, but not any of the new oblique
parameters. They include $\rho_*(0)$, $M_W/M_Z$, and various ratios 
and asymmetries measured on the $Z$-mass. 
Since the SM and the BSM contributions
add linearly and since \cite{one}, for $m_t=175$ GeV and $m_H=100$ GeV,
\be
S_{SM}=0.60,\ \ T_{SM}=0.79,\ \ U_{SM}=0.95,
\ee
we constrain the BSM contributions, denoted by a tilde overhead, to be
\be
\ts=-0.04\pm 0.26,\ \ \tt=-0.04\pm 0.31,\ \ \tu=-0.63\pm 0.61.
\ee
Within $1\sigma$ error, these are perfectly consistent with zero (no
observable BSM physics). The fact that the allowed ranges for $\ts$,
$\tt$ and $\tu$ are nearly the same as those obtained using the linear
approximation proves the robustness of the linear approximation. 

We now proceed to consider V,W and X. First note that the SM values of
V,W and X are of the order of $10^{-3}$. More precisely, for $m_t=175$ GeV
and $m_H=100$ GeV, $V_{SM}=-0.004$, $W_{SM}=-0.003$ and $X_{SM}=0.001$.
Thus the constraints obtained on these new oblique parameters are essentially
those on $\tv$, $\tw$ and $\tx$.
The constraint on V can be obtained from the total decay width of $Z$, and
comes out to be
\be
\tv=0.30\pm 0.38.
\ee
Similarly, the constraint on W can be obtained from the total decay width
of the $W$-boson, or its partial width to leptons. One finds that
\be
\tw=0.11\pm 4.70.
\ee
Because of the greater inaccuracy in
the present data on the $W$, relative to
those on the $Z$, this constraint on W is one order of magnitude weaker as
compared to that on V. 
Coming to X, we constrain it from observables at 
$q^2=0$ as follows:
\be
\tx =  0.38\pm 0.59.
\ee

As expected, all of these new oblique parameters are compatible with zero,
and central values are of the same order as those of $\ts$, $\tt$
and $\tu$. 

In the $\star$ scheme explicit expressions for the oblique parameters can 
be written in terms of observable quantities. For S,T,U, such expressions 
were already given by Peskin and Takeuchi
\cite{pt}. We write those for V,W,X below.
\be
V=162.0+55.0\G_Z-7.8R_l-137.0\rho_*(0),
\ee
\be
W=582.0-15.52R_l+9.89\rho_*(0)-461.72{M_W\over M_Z}+0.137Q_W({ }^{133}_
{~55}Cs)+66.34\G_W,
\ee 
\be
X=163.0-7.90R_l+46.90\rho_*(0)+0.62Q_W({ }^{133}_{~55}Cs).
\ee
Here $R_l\equiv\G(Z\to {\rm hadrons})/\G(Z\to\ell\bar\ell)$.
Of course, V,W and X can be written in terms of other observables too, but we
have chosen the set with the least experimental errors and hence maximum 
sensitivity to any possible BSM contributions.

\bigskip\bigskip

\centerline{\large VI. {\bf Summary and Discussion}}

\bigskip
 
Three salient features of our work can be summarized as follows.

\begin{enumerate}

\item[{$\bullet$}] 
Electroweak oblique parameters S,T,U,V,W and X can be consistently
defined at the one-loop level with the help of $\p$, $\p'$ and $\p''$
functions and the $q^2$-derivative of the
$\p''$ function --- without any $q^2$ expansion approximation.
The symmetry content of S,T and U are manifest in our definitions.

\item[{$\bullet$}] 
The new oblique parameters V and W appear only in the wavefunction 
renormalizations of the $Z$ and the $W$ boson respectively, and through
them, in the expressions for the total and partial decay widths of these
bosons. X appears only in $s_*^2(0)$ and through it, in other   
observables (such as Moller asymmetry
\cite{czarnecki}) at $q^2=0$. The physics contents of these new parameters 
are quite distinct.

\item[{$\bullet$}] 
The constraints on the new oblique parameters are as follows: $V=
0.30\pm0.38$, $W=0.11\pm4.70$, $X=0.38\pm 0.59$, and there is
practically no difference here between those with tilde and those 
without.

\end{enumerate}

What is really the new element in our work? We have given definitions
of oblique parameters with correct symmetry properties.  
We have also shown that ratios 
of LEP and Tevatron observables, that are independent of the renormalization
constants $Z_{W,Z}$, involve only S,T and U, and give no information on
the new oblique parameters V,W and X. We have further considered LEP and 
Tevatron observables that involve $Z_Z$ and $Z_W$; we have been able to 
relate $Z_Z$ and $Z_W$ to the oblique parameters and to derive numerical
bounds on V and W. On the other hand, constraints on X follow from low-energy
measurements. We have also shown that the linear approximation of Peskin 
and Takeuchi is, in fact, quite robust. This is because
the bounds on S,T,U depend very little on whether one uses
the linear approximation or not. It is nonetheless interesting
to explore the new parameters V,W and X as they can be significant if
there is relatively light BSM physics. With that objective, we have
expressed V,W and X as functions of those experimentally measured 
quantities which are being measured most accurately. These can provide
suitable precision probes for BSM physics with respect to better
data in future. Our stress in this paper, however,
has been more on formal issues concerning oblique parameters and less on
precision testing. The numbers quoted are, therefore, correct only within
the framework of the $\star$ scheme; one-loop non-oblique corrections 
as well as significant two-loop effects involving the top quark have not
been taken into account. A precision analysis should include those 
non-oblique correction, in view of the highly accurate results on the 
$Z$-peak. Nevertheless, the general definitions of 
oblique parameters given in this work as well as the accompanying 
discussion of the expansion approximation together with the suggestive
evidence for the robustness of the linear expansion will hopefully be
useful in clarifying the broad picture.

%\newpage
%\bigskip\bigskip

\centerline {\bf Acknowledgements}

This work originated in the workshop WHEPP-3 organized by the Institute
of Mathematical Sciences, Madras, under the sponsorship of the S.N. Bose
National Centre for Basic Sciences. PR acknowledges the hospitality
of Oxford University where he did part of this work under the
Royal Society/Indian National Science Academy exchange programme.
We acknowledge discussions with F. Caravaglios, M. Einhorn, S. King,
P. Langacker and M. Peskin. We are especially indebted to 
Sunanda Banerjee for his assistance with experimental data.

\newpage
\appendix
\setcounter{equation}{0}
\renewcommand{\theequation}{A.{\arabic{equation}}}
\centerline{\large\bf Appendix}
\bigskip

In this appendix, we give the numerical expressions for observables measured 
at the $Z$- (or the $W$) peak, and at $q^2=0$. First we quote those
observables which can be expressed as functions of S,T and U only.
As already noted in the text, these expressions are only valid within the
framework of the $\star$ scheme. Thus, we do not qoute any error margins 
in these expressions, as the precision of the $\star$ scheme can be 
questioned where two-loop effects, threshold effects and other non-oblique
corrections are significant.

\bea
s_*^2(\mzs)&=&0.231[1+0.014S-0.010T],\\
\rho_*(0)&=&1+7.29\t\mt,\\
R_l&\equiv&\G_h/\G_{\ell\bar\ell}=20.787[1-2.78\t\ms+1.97\t\mt],\\
R_b&\equiv&\G_b/\G_h=0.216[1+0.54\t\ms-0.41\t\mt],\\
R_c&\equiv&\G_c/\G_h=0.171[1-1.17\t\ms+0.79\t\mt],\\
\sigma_h^0&=&41.464[1+0.59\t\ms-0.25\t\mt],\\
M_W/M_Z&=&0.877[1-1.19\t\ms+3.68\t\mt+4.06\t\miu],\nonumber\\
& &\\
A_{LR}&=&0.148[1-0.175S+0.122T]=-P_{\tau},\\
A^l_{FB}&=&0.75(A_{LR})^2=0.016[1-0.350S+0.245T],\\
A^b_{FB}&=&0.104[1-0.177S+0.124T],\\
A^c_{FB}&=&0.073[1-0.192S+0.134T].
\eea

Next, those expressions which contain V,W and X, apart from S,T and U,
are listed.

\bea
s_*^2(0)& =& 0.2312[1+0.014 S-0.010 T-0.032 X],\\
\G (Z\r \nu\bar\nu)& =&0.166[1-0.35\t 10^{-3} S+7.10\t 10^{-3}
T+7.29\t 10^{-3} V],\nonumber\\
& &\\
\G (Z\r \ell\bar\ell)& =&0.084[1-1.65\t 10^{-3} S+8.45\t 10^{-3}
T+7.29\t 10^{-3} V],\nonumber\\
& &\\
\G (Z\r u\bar u)& =&\G (Z\r c\bar c)\nonumber\\
&=&
0.297[1-5.6\t 10^{-3} S+11.2\t 10^{-3}
T+7.29\t 10^{-3} V],\nonumber\\
& &\\
\G (Z\r d\bar d)& =&\G (Z\r s\bar s)\nonumber\\
&=& 
0.383[1-3.8\t 10^{-3} S+10.0\t 10^{-3}
T+7.29\t 10^{-3} V],\nonumber\\
& &\\
\G (Z\r b\bar b)& =&0.375[1-3.9 \t 10^{-3} S+10.0\t 10^{-3}
T+7.29\t 10^{-3} V],\nonumber\\
& &\\
\G _h& =&2\G _u+2\G_d+\G_b\nonumber\\
&=&1.735[1-4.4 \t 10^{-3} S+10.4\t 10^{-3}
T+7.29\t 10^{-3} V],\nonumber\\
& &\\
\G _Z& =&\G_{\nu\bar\nu}+\G_{\ell\bar\ell}+\G_h\nonumber\\
&=&2.484[1-3.3 \t 10^{-3} S+9.6 \t 10^{-3}
T+7.29\t 10^{-3} V],\nonumber\\
& &\\
\G (W\r \ell\bar\nu)& =&0.224[1-0.010S+0.015T+0.012U \nonumber\\
& & ~~~~~~~+7.29\t 10^{-3} W -1.6\t 10^{-4} X],\\
\G _W& =&2.065[1-0.010S+0.015T+0.012U \nonumber\\
& & ~~~~~~~+ 7.29\t 10^{-3} W-1.6\t 10^{-4} X],\\
g_L^2& =&0.298[1-0.83\t 10^{-2} S+2.04\t 10^{-2}
T+1.82\t 10^{-2} X],\nonumber\\
& &\\
g_R^2& =&0.030[1+2.88\t 10^{-2} S-0.56\t 10^{-2}
T-6.32\t 10^{-2} X],\nonumber\\
& &\\
R_{\nu}& =&0.310[1-6.94\t 10^{-3} S+19.45\t 10^{-3}
T+15.19\t 10^{-3} X],\nonumber\\
& &\\
R_{\bar\nu}& =&0.378[1-0.45\t 10^{-3} S+14.92\t 10^{-3}
T+0.98 \t 10^{-3} X],\nonumber\\
& &\\
Q_W({ }^{133}_{~55}Cs)& =&-73.855[1+9.9 \t 10^{-3} S+0.4 \t 10^{-3}
T-21.7\t 10^{-3} X].\nonumber\\
& &
\eea 

As already discussed in the text, we have used the quadratic approximation 
\cite{seven} in deriving (A.20) and (A.21), since these expressions
involve $s_*^2(\mws)$ which cannot be evaluated without such an 
approximation.  At any arbitrary $q^2$, $s_*^2$ can then be written as
\be
s_*^2(q^2)=\sz+{\a\over c^2-s^2}\Big[{1\over 4}S-c^2s^2T\Big]-\a\Big(1-
{q^2\over\mzs}\Big)X.
\ee
(A.27) assumes that
$\p'_{\g Z}$ has a constant slope from $q^2=0$ to $q^2=\mzs$ --- which is 
the quadratic approximation. At $q^2=0$ or $q^2=\mzs$, no such approximation
is needed; the expression is exact. However, at $q^2=\mws$, the
quadratic-approximated (A.27) yields
\be
s_*^2(\mws)=\sz+{\a\over c^2-s^2}\Big[{1\over 4}S-c^2s^2T\Big]-\a s^2X.
\ee

\end{document}